\documentclass{an}
\usepackage{graphicx}
\usepackage[authoryear]{natbib}

\bibpunct{(}{)}{,}{a}{}{;}
\begin{document}
\Pagespan{1}{}% Document's page range. 
\Yearpublication{}%
\Yearsubmission{2007}%
\Month{}%   
\Volume{}%  
\Issue{}% 
% \DOI{This.is/not.aDOI}% 

\title{Radio Wavelength Transients: Current and Emerging Prospects}
\author{J.~Lazio (NRL)}
\institute{ 
Naval Research Laboratory, 4555 Overlook Ave.~\hbox{SW}, Washington,
\hbox{DC}, 20375  USA} 
 
\titlerunning{Radio Wavelength Transients} 
\authorrunning{Lazio} 

\received{2007 August 31} 
\accepted{} 
\publonline{later} 
 

\abstract{
Known classes of radio wavelength transients range from the
nearby---stellar flares and radio pulsars---to the distant
Universe---$\gamma$-ray burst afterglows.  Hypothesized classes of
radio transients include analogs of known objects, e.g., extrasolar
planets emitting Jovian-like radio bursts and giant-pulse emitting
pulsars in other galaxies, to the exotic, prompt emission from
$\gamma$-ray bursts, evaporating black holes, and transmitters from
other civilizations.  A number of instruments and facilities are
either under construction or in early observational stages and are
slated to become available in the next few years.  With a combination
of wide fields of view and wavelength agility, the detection and study
of radio transients will improve immensely.}

\maketitle

\section{Radio Transients}\label{sec:jl.science}

Transient radio sources are necessarily compact and usually are the
locations of explosive or dynamic events, therefore offering unique
opportunities for probing fundamental physics and astrophysics.  In
addition, short-duration transients are powerful probes of intervening
media owing to dispersion, scattering and Faraday rotation that modify
the signals.  Searches for radio transients have a long history, and a
wide variety of radio transients are known, ranging from extremely
nearby to cosmological distances.  In addition, motivated by analogy
to known objects or applying known physics, there are a number of
classes of hypothesized classes of transients.

\begin{description}
\item[\textbf{Ultra-high Energy Particles}:]
Short-duration ($\sim 1$ ns) pulses have been observed from the impact
of ultra-high energy particles on the Earth's atmosphere
\citep{w01}.  High-energy neutrinos impacting the lunar regolith
should also produce radio pulses, though searches to date have been
unsuccessful \citep{heo96}.  Detection of such particles can place
significant constraints on the efficiency of cosmic accelerators.

\item[\textbf{The Sun}:]
Type~II and~III solar bursts are detected regularly
\citep{mkcasms96,pscdm88} and solar flares can also be detected
\citep{wksysk03}.  As such, the Sun offers a nearby site to study
particle acceleration in detail, particularly when observations are
combined with optical, ultraviolet, and X-ray observations.

\item[\textbf{Planets}:]
Jupiter has long been known to produce decameter bursts
\citep{bf55}, and all of the planets with strong magnetic fields
(Earth, Jupiter, Saturn, Uranus, and Neptune) produce radio radiation,
though not always above the Earth's ionospheric cutoff.  At least one
of the known extrasolar planets also appears to have a strong magnetic
field \citep{swbgk05}.  By analogy to the solar system planets,
\cite{zqr+97} and \cite{fdz99} suggest that extrasolar giant planets
also produce bursty emission.  Detection of radio emission from
extrasolar planets would constitute \emph{direct} detection, in
contrast to the largely indirect detections of the known extrasolar
planets\footnote{
The first extrasolar planets were discovered
using radio observations of the pulsar PSR~B1257$+$12, though any
direct radio emission from these pulsar planets is not likely to be
detectable.}
obtained from the reflex motions of their host stars.

\item[\textbf{Brown Dwarfs}:]
Radio flares have been detected from BD~LP944$-$20
\citep{bergeretal01} and TVLM~513-46546 \citep{hbl+07}, among others
\citep{b02}.  The flares are thought to originate in magnetic activity
on the surfaces of the brown dwarfs, but some objects have radio
emission that deviates from the expected radio--X-ray correlation
observed for stars.

\item[\textbf{Flare Stars}:]
Radio flares attributed to particle acceleration from magnetic field
activity are observed from various active stars and star systems at
frequencies of order 1~GHz \citep{g02}.

\item[\textbf{Pulsar Giant Pulses}:]
Some pulsars emit ``giant''\hfil\linebreak
pulses, with strengths that are 1000 times the mean pulse intensity.
In the case of the Crab pulsar, some giant pulses outshine the
entire Crab Nebula, on time scales of roughly 100~$\mu$s \citep{hr75},
corresponding to a brightness temperature of~$10^{31}$~\hbox{K};
recently, \cite{hkwe03} observed ``nano-giant'' pulses having
durations of only 2~ns and implied brightness temperatures of
10$^{38}$~K, by far the most luminous emission from any astronomical
object.  For many years, this phenomenon was thought to be unique to
the Crab pulsar, but giant pulses have since been detected from other
pulsars \citep{cstt96,jr03}.

\item[\textbf{Transient Pulsars}:]
%\cite{n99} discovered a pulsar using a single pulse search during a
%re-analysis of a Galactic plane survey; this pulsar (J1918$+$08) does
%not emit giant pulses but is a normal, slow pulsar which fortuitously
%emitted one strong pulse during the search observations.
\cite{klojll06} have recognized a class of pulsars that
produce pulses only a small fraction of the time.  For instance, PSR\hfil\linebreak
B1931$+$24 is detectable less than 10\% of the time.  Single-pulse
searches of pulsar surveys also have resulted in the discovery of
``rotating radio transients,'' neutron stars whose emission is so
sporadic that they are not detectable in standard Fourier domain
searches \citep{mll+06}.

\item[\textbf{X-ray Binaries}:]
Large radio flares, with peak flux\hfil\linebreak
densities being factors of~10--100 larger
than quiescence, have long been known from X-binaries such as
Cyg~X-3 \citep{wgjffs95}.  Searches for short-duration, single
pulses from the X-ray binaries Sco~X-1 and Cyg~X-1 have been
unsuccessful \citep{thh72}.
% < 10 Jy in 5 ms pulses

\item[\textbf{Soft $\gamma$-ray Repeaters}:]
\cite{vl87}\hfil\linebreak
searched, unsuccessfully,  for radio pulses from the soft
$\gamma$-ray repeater SGR~0526$-$66.

\item[\textbf{Masers}:]
The emission from OH masers can vary on timescales of hundreds of
seconds and be detected as long-duration radio bursts \citep{cb85,y86}.

\item[\textbf{Active Galactic Nuclei}:]
Active galactic nuclei outbursts, likely due to propagation of shocks
in relativistic jets, are observed at millimeter and centimeter
wavelengths \citep{aalh85,l94}.

\item[\textbf{Intraday Variability}:]
Intraday variability results\hfil\linebreak
from interstellar scintillation of extremely compact components ($\sim
10$~$\mu$as) in extragalactic sources.  The typical modulation
amplitude is a few percent and is both wavelength and direction
dependent, but occasional rare sources display much larger amplitude
modulations on timescales of hours to days \citep{k-cjwtr01,ljbk-cmrt03}.

\item[\textbf{Radio Supernovae}:]
\cite{cn71}\hfil\linebreak
predicted that a supernova explosion should produce a large, broadband
radio pulse ($< 1$~s), though no such pulses have yet been
detected\hfil\linebreak
\citep{hm74,kardashevetal77}.

\item[\textbf{$\gamma$-ray Bursts}:]
\cite{cmmcis81},\hfil\linebreak
\cite{ismm82}, and \cite{alv89} detected dispersed
radio pulses but found no convincing associations with gamma-ray burst
sources.
%\cite{b99} found a dispersed radio pulse apparently
%coincident with GRB~980329; however, it was narrowband, which has led
%to it being interpreted as due to terrestrial interference.
Various searches for radio pulses associated with gamma-ray bursts
(including precursor pulses) have been conducted
\citep{kgwwp94,kgwwp95}.  More recently, \cite{uk00} and \cite{sw02}
have predicted that gamma-ray bursts should have associated prompt
emission, most likely below 100 MHz.

\item[\textbf{Gravitational Wave Sources}:]
In a search for radio counterparts to the gravitational pulses
reported by \cite{w69}, both \cite{hr73} and \cite{ehm74} detected
excesses of radio pulses from the direction of the Galactic center,
but did not believe them to be correlated with the gravitational
pulses.  More recently, \cite{hl01} have predicted that inspiraling
neutron star-neutron star binaries could produce radio precursors to
the expected gravitational wave signature.

\item[\textbf{Black Holes}:]
\cite{r77} predicted that annihilating black holes might produce radio
bursts, though searches have not yet found a convincing signal\hfil\linebreak
\citep{oes78,pt79}.

\item[\textbf{Extraterrestrial Transmitters}:]
While no such examples are known of this class, many searches for
extraterrestrial intelligence (SETI) have found non-repeating signals
that are otherwise consistent with the expected signal from an
extraterrestrial transmitter.  \cite{cls97} discuss how
extraterrestrial transmitters could appear to be transient, even if
intrinsically steady.
\end{description}

\section{Emerging Instruments}\label{sec:jl.telescope}

Here we describe briefly a number of the facilities that are either
under construction or in the design and development phases.  In
essentially all cases, transients form a key portion of the science
case of these instruments.  An informal collaboration has been started
among various team members to assess how to identify and communicate
radio transients; future work will include how to interface to
facilities at other wavelengths.

\subsection{Allen Telescope Array (ATA)}\label{sec:jl.telescope.ata}

The ATA is a project designed to develop a survey-oriented radio
telescope, with a strong emphasis on surveys for transients.  The ATA
will cover the wavelength range 3 to~60~cm (0.5--11~GHz)
instantaneously, with multiple, simultaneous spectral windows within
this range.  The angular resolution will be modest ($\approx
1^\prime$), but with a large instantenous field of view ($\approx 4$
deg.${}^2$ at~20~cm) and a drive system designed to access any region
of the sky quickly ($< 1$~min.).  Currently, 42 antennas are deployed
and operating; if sufficient funding can be identified, the telescope
will be expanded to~350 antennas.

\subsection{Expanded Very Large Array
	(EVLA)}\label{sec:jl.telescope.evla}

The EVLA is a project to construct an entirely new telescope on the
foundation of the Very Large Array.  Replacing the original
electronics packages with modern digital electronics, the EVLA will
cover the entire wavelength band from~0.7 to~30~cm (1--50~GHz), with
modest additional coverage of longer wavelengths ($\sim 1$~m).  The
expanded spectral converage will enable (continuum) sensitivities
approaching 1~$\mu$Jy.  The first antennas are entering service now,
and construction is on-schedule for completion in~2012.

\subsection{Long Wavelength Array (LWA)}\label{sec:jl.telescope.lwa}

The LWA will observe between~3.75 and~15~m wavelength (20--80~MHz),
with the goal of obtaining arcsecond resolution over this entire
wavelength range.  Composed of approximately 50 ``stations'' of
dual-\hfil\linebreak 
polarization dipoles, the LWA will be located in the
\hbox{U}.\hbox{S}.\ state of New Mexico.  The dipoles in each station
will be phased together to form a beam that can be electronically
steered, on rapid time scales, to any point in the sky.  The signals
from the stations will be combined interferometrically to form a
synthesized beam and obtain the arcsecond resolution.  The total
number of dipoles anticipated is of order 10\,000, which will provide
a sensitivity of~1--10~mJy.  Currently deployed, at the center of the
\hbox{VLA}, is the prototype, 16-element Long Wavelength Demonstrator
Array (LWDA), which is already being used to test methods for all-sky
transient monitoring at these wavelengths.

\subsection{Low Frequency Array (LOFAR)}\label{sec:jl.telescope.lofar}

LOFAR will observe in two bands, 3.3 to~10~m (30--80~MHz) and~1.2
to~2.5~m (110--240~MHz), employing two different designs.  The low
band (30--80~MHz) is designed much like the \hbox{LWA}, with dipole
stations.  The stations will be distributed throughout the northern
part of the Netherlands, with extensions into other countries in
Europe.  In the high band (110--240~MHz), the basic element is a
``tile,'' composed of~16 dipoles (4 $\times$ 4) phased together.
Thus, a full LOFAR station is composed of a cluster of low-band
dipoles and high-band tiles spread throughout a 50~m area.  In the
lower band, LOFAR will have a sensitivity comparable to the
\hbox{LWA}; in the higher band, it will have a sensitivity of better
than 0.1~mJy.  Currently deployed are 96 low-band dipoles, arrayed
throughout what will be the first station, located in the northeast of
the Netherlands.  Initial all-sky images have been produced, and a
transient pipeline is being tested.

\subsection{Murchison Wide-field Array
	(MWA)}\label{sec:jl.telescope.mwa}

The MWA will observe between~1 and~3.75~m (80--300~MHz).  Similar to
the LOFAR high-band elements, the basic element of the MWA is a
16-dipole tile.  Currently deployed in Western Australia are 32
tiles, en route to the full deployment of~500 tiles.  Observations
have begun with these three tiles, with giant pulses from the Crab
pulsar detected \citep{bwk+07}.  The full MWA will have a sensitivity
of better than 1~mJy and will cover about 1.5~km, providing an angular
resolution of several arcminutes.

\subsection{Other Instruments}\label{sec:jl.telescope.other}

Funding has just been announced for two additional instruments.  In
Australia, the Australian SKA Pathfinder (ASKAP) will be constructed
in the state of Western Australia, in one of the most ``radio quiet''
(or radio dark) locations on the planet.  Design plans are for
approximately twenty (20), 15-m diameter antennas operating at a
wavelength around~21~cm and equipped with a phased-array feed or
multi-pixel detector.  Optimized for surveys, including radio surveys,
the telescope will be an ideal transient detector.

In South Africa, the Karoo Array Telescope\hfil\linebreak
(MeerKAT) will be located in the northern province of Karoo, also an
exceptionally radio quiet location.  Design plans are for as many as
80 antennas, likely to be 12~meters in diameter, initially equipped
with only single-pixel detectors.  Its coverage will include
wavelengths around 21~cm, but may extend to as short as 3~cm.

\acknowledgements
I thank G.~Bower, J.~Cordes, J.~Kasper, C.~Law, and Y.~Philstr\"om for helpful
comments.  Basic research in radio astronomy at the NRL is supported
by 6.1 Base funding.

%\subsection{Square Kilometre Array (SKA)}\label{sec:jl.telescope.ska}

\end{document}